# An Insight View of Kernel Visual Debugger in System Boot up

Mohamed Farag

Department of Computer Science, Maharishi University of Management, USA
`mfarrag@freebsd.org`

## ABSTRACT

*For many years, developers could not figure out the mystery of OS kernels. The main source of this mystery is the interaction between operating systems and hardware while system's boot up and kernel initialization. In addition, many operating system kernels differ in their behavior toward many situations. For instance, kernels act differently in racing conditions, kernel initialization and process scheduling. For such operations, kernel debuggers were designed to help in tracing kernel behavior and solving many kernel bugs. The importance of kernel debuggers is not limited to kernel code tracing but also, they can be used in verification and performance comparisons. However, developers had to be aware of debugger commands thus introducing some difficulties to non-expert programmers. Later, several visual kernel debuggers were presented to make it easier for programmers to trace their kernel code and analyze kernel behavior. Nowadays, several kernel debuggers exist for solving this mystery but only very few support line-by-line debugging at run-time. In this paper, a generic approach for operating system source code debugging in graphical mode with line-by-line tracing support is proposed. In the context of this approach, system boot up and evaluation of two operating system schedulers from several points of views will be discussed.*

## 1. INTRODUCTION

A kernel is a central component of an operating system. It acts as an interface between the user applications and the hardware. The sole aim of the kernel is to manage the communication between the software (user level applications) and the hardware (CPU, disk memory etc). The main tasks of the kernel are: Process Management, Device Management, Memory Management, Interrupt Handling, I/O Communication and File System. Debugging is a methodical process of finding and reducing the number of bugs, or defects, in a computer program or a piece of electronic hardware, thus making it behaves as expected. As software and electronic systems have become generally more complex, the various common debugging techniques have expanded with more methods to detect anomalies, assess impact, and schedule software patches or full updates to a system. An OS kernel debugger is a debugger presented insome kernels for ease debugging and kernel development. There are several techniques to implement kernel debugging such as printing debugging, remote debugging, post-mortem debugging[1], delta debugging[2] and Saff Squeez[3]. Remote debugging is the process of debugging a program running on a system different than the debugger machine. In remote debugging, a debugger connects to a remote system over a network. Once connected, the debugger can control the execution of the program on the remote system and retrieve information about its execution. Remote debugging follows client/server architecture with TCP/IP or COM for communication between machines.The role of the client machine is to debug the code existing on the server but server machine's application should have enabled debugging support to permit another machine to debug its code. In addition, a copy of the application running on the remote hardware should be kept in the client side to be able to view it. This copy should have the same debug information as the original. If the remote application is using another target(i.e. different hardware architecture), client debugger should also support target's platform.

## 2. RELATED WORK

Several attempts were made for kernel debugging. Several visual kernel debuggers were designed for this purpose such as DDD and WinDbg. GNU DDD is a graphical front-end for command-line debuggers such as GDB, DBX, WDB, Ladebug, JDB, XDB, the Perl debugger, the bash debugger bashdb, the GNU Make debugger remake, or the Python debugger pydb. Besides ``usual" front-end features such as viewing source texts, DDD has become famous through its interactive graphical data display, where data structures are displayed as graphs. WinDbg is a multi- purposed debugger for Microsoft Windows, distributed on the web by Microsoft. It can be used to debug user mode applications, drivers, and the operating system itself in kernel mode. It is a GUI application, but has little in common with the more well-known, but less powerful, Visual Studio Debugger[5].

## 3. PROBLEM DESCRIPTION & SUGGESTED SOLUTION

Existing visual kernel debuggers have some problems. Many visual debuggers supporting kernel debugging are front end views for kernel debuggers so some of them are not fully compatible with the debugger itself[6]. For example, DDD debugger which is a front end for GNU GDB but it doesn't work with old editions of GDB. In addition, some debuggers are platform dependant such as WinDbg which is Windows-specific debugger. On the other hand, some debuggers don't provide assembly level debug along with C debugging rather than visualizing the code[4]. Furthermore, debuggers should use at least two machines to debug the code remotely hence making it costly to test such machines. According to the mentioned in the last paragraph, there's a solution satisfying the following criteria:

1. Platform-Independent.
2. Portable.
3. Supporting the debug of different programming languages at the same run (C and Assembly).
4. High compatibility between the front end and the debugger itself.
5. User friendly.
6. Low Cost.
7. Open Source with appropriate license (GPL and EPL).

These criteria can be satisfied using tools reconfiguration and there's no need to develop new tool or write code for that.The solution we are providing is an enhanced version ofremote debugging.However, Fig 1 shows typical remote debugging environment while the topology followed in this paper is the one provided in Fig 2.

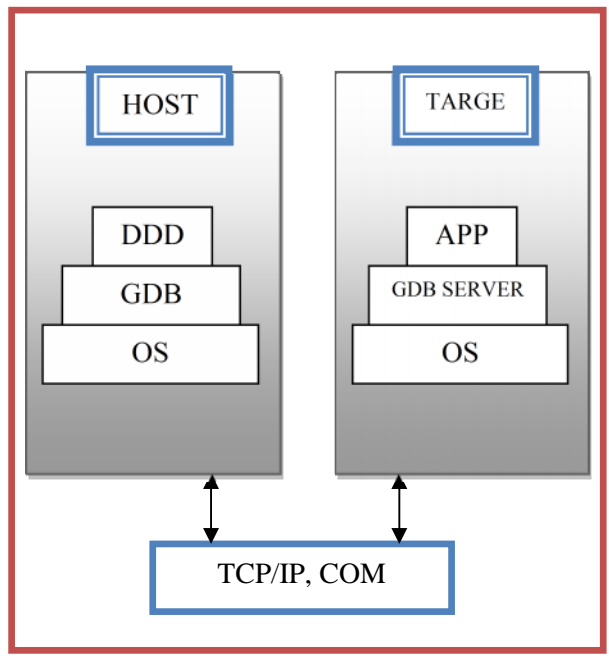

Fig 1 Remote Debugging Topology

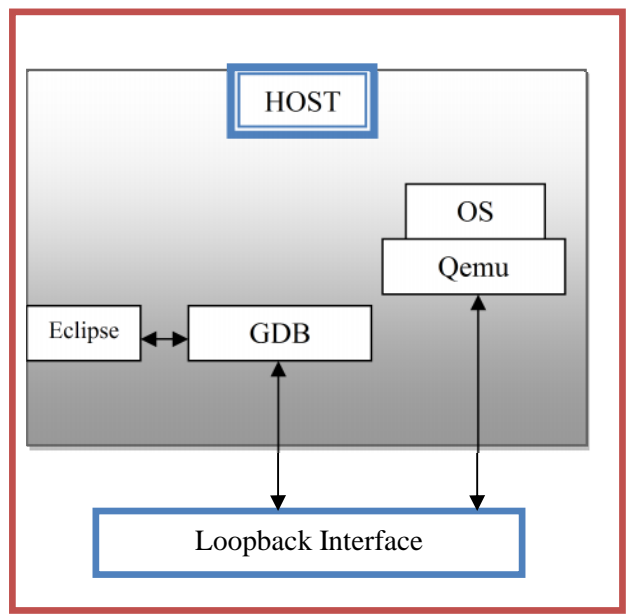

Fig 2 Enhanced Remote Debugging Topology

In the suggested approach, different metrics were in consideration according to the following points:

1.  Take the advantage of Eclipse CDT and Qemu as portable, platform independent, Open Source and reliable tools supporting this technique under several platforms.
2.  Eclipse support different debuggers and can combine many of them under the same run.
3.   Eclipse has its own GDB plugin which guarantees high compatibility.
4.  Only one machine is used for lower cost.

In suggested approach, Eclipse CDT, which supports C & assembly languages, connects to GDB Server for debugging. GDB Server used TCP/IP protocol stack with tricky loopback interface to connect to Qemu. In separate thread of execution, Qemu is pointed to target kernel image and controlled by GDB Server commands to trace kernel execution.

## 4. METHODOLOGY

The main idea behind this implementation is to command Qemu to point to the kernel boot initializer. Behind the scene, boot initializer will initialize the kernel and start it. In the same time, GDB Server debugger is tracking the kernel image itself to provide connection point between Eclipse and Qemu through recognizing line of code being executed and commanding Qemu. Let's consider linux kernel debugging to clarify the state of the art in this idea,. If linux kernel is compiled, kernel boot initializer will be stored in arch/*machine_arch*/boot/bzimage under the kernel compilation directory. In addition, the kernel image will be co-located in the parent kernel directory under the name vmlinux. The following figure describes this scenario in detail. Please consider the numbers above the arrows to indicate the order of steps.

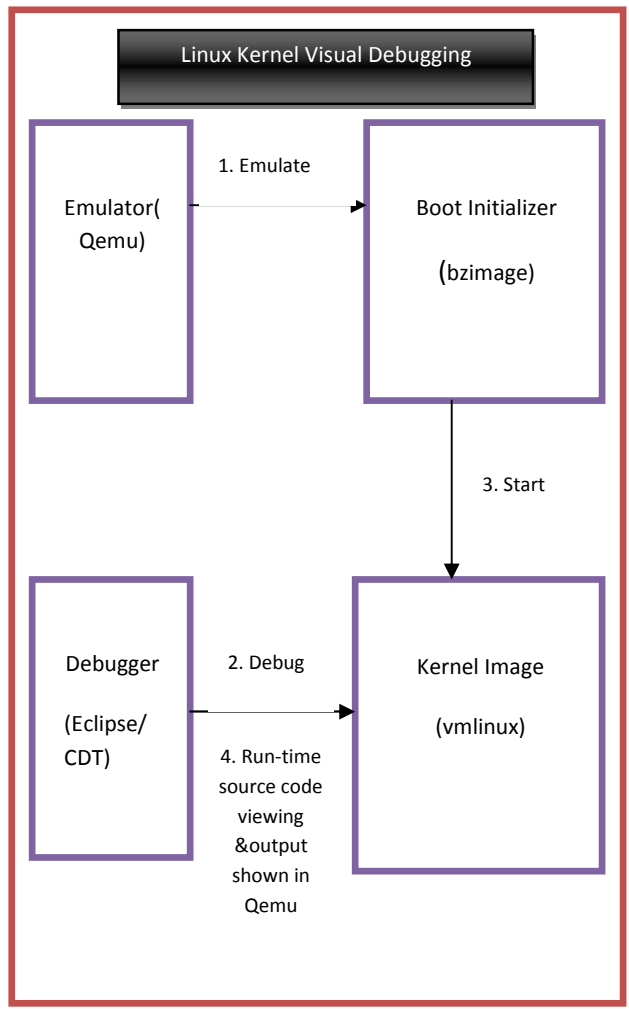

Fig 3 Run-time Debugging Communication Diagram

The following figure shows an eclipse screen shot at the scheduler initialization function "sched_init()".

Fig 4 Eclipse Debugger screenshot

## 5. CASE STUDY (FREEBSD OS BOOT)

There's no doubt that system scheduler is the most important part of operating system due to its performance effect according to preemption or non-preemption rules. One of the great debugging mysteries is the system's boot-up process which describes how kernel could be initialized and turned on to an active state. Nowadays, debuggers provide huge capabilities such as performance measurements and code verification. One of the approaches used for performance measurements is Line Of Code (LOC) approach. Usually, number of line of code will be multiplied by the average time to execute one command (Typically, 10ns) resulting the total average time. Next, this technique was used to analyze FreeBSD Boot-up and to compare the performance of two common FreeBSD schedulers. The first one is ULE Scheduler and the second one is 4.3 BSD Scheduler. Notice that another platform is used in this case study to show how compatible is this approach with other platforms. FreeBSD 8.0 Machine on 32-bit is used for this experiment. The first output of the program was control flow of FreeBSD booting process.

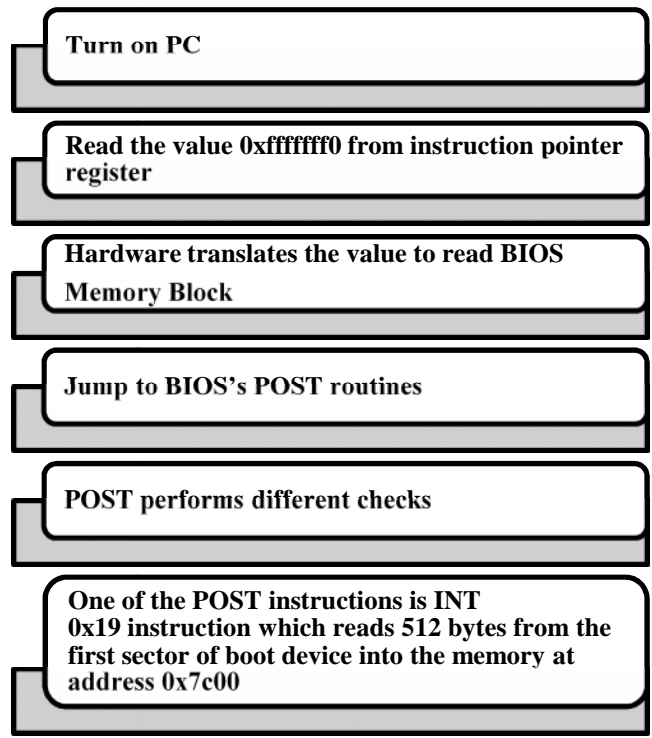

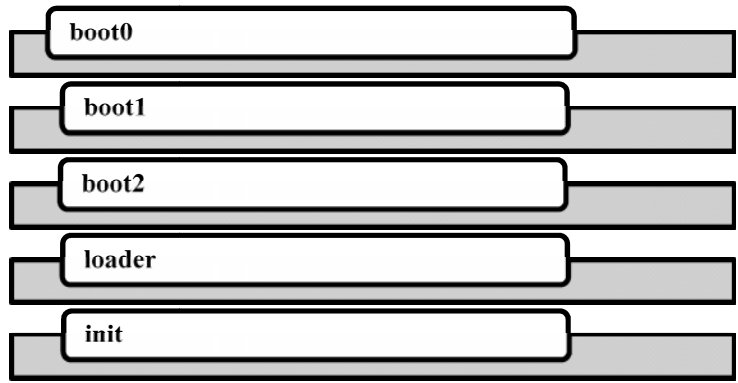

Fig 5 Control Flow for FreeBSD 8.0

An important advantage of kernel source level debugging is to detect the control flow diagram. On the other hand, it can be used to specify function responsibilities and how they relate to the overall execution. Next, all the boot functions are described in the following steps:

1. boot0

| Location | sys/boot/i386/boot0/boot0.S |
|---|---|
| **Memory** | INT 0x19 instruction loads an MBR, i.e. the boot0 content, into the memory at address 0x7c00 |
| **MBR Structure** | Starting from 0x1be, called the *partition table,* It has 4 records of 16 bytes each, called *partition record* |
| **Partition Record Structure** | • 1-byte file system type.<br>• 1-byte bootable flag.<br>• 6 byte descriptor in CHS format.<br>• 8 byte descriptor in LBA format. |

Table 1.0 boot0 properties

2. boot2

| File Location | sys/boot/i386/boot2/boot2.c |
|---|---|
| **Functionality** | • It scans the harddisk.<br>• knowing about the filesystem structure<br>• finds the file /boot/loader<br>• reads it into memory using a BIOS service<br>• passes the execution to the loader's entry point.<br>• boot2 prompts for user input so the loader can be booted from different disk, unit, slice and partition<br>• passes the execution to BTX |

Table 2.0 boot2 properties

3. Loader

| File Location | sys/boot/i386/boot/loader |
|---|---|
| Notes | • kernel bootstrapping final stage<br>• When the kernel is loaded into memory, it is being called by the loader |
| Functionality | • The main task for the loader is to boot the kernel.<br>• It provides a scripting language that can be used to automate tasks, do pre-configuration or assist in recovery procedures. |

Table 3.0 Loader properties

4. init386()

| File Location | sys/i386/i386/machdep.c |
|---|---|
| Functionality | • Initialize the kernel tunable parameters, passed from the bootstrapping program.<br>• Prepare the GDT.<br>• Prepare the IDT.<br>• Initialize the system console.<br>• Initialize the DDB, if it is compiled into kernel.<br>• Initialize the TSS.<br>• Prepare the LDT.<br>• Set up proc0's pcb |

Table 4.0 init386 Properties

After complete analysis of system component responsibilities, this approach was used in more effective way to compare the execution time of two common schedulers in FreeBSD. The following results show that 4.3BSD scheduler is much more efficient than ULE scheduler through the installation of Apache port. Time units are in seconds but the most important are the differences in the results. On Pentium-4 Machine, real time statistics are presented in the following table:

| Concurrent Processes | ULE | ULE Stddev | BSD | BSD Stddev | faster |
|---|---|---|---|---|---|
| 2 | 2371.9 | 1.212 | 2346.29 | 1.89 | 4BSD |
| 4 | 2007.8 | 2.58 | 1999 | 0.68 | 4BSD |

Table 5.0 Real Time Statistics for Schedulers

User Time statistics are presented in the following table:

| Concurrent Processes | ULE User | ULE Stddev | BSD User | BSD Stddev | faster |
|---|---|---|---|---|---|
| 2 | 2251.9 | 5.3 | 2221.6 | 2.1 | 4BSD |
| 4 | 2499.9 | 2.74 | 2416.13 | 2.9 | 4BSD |

Table 6.0 User Time Statistics

System Time statistics are calculated in the following table:

| Concurrent Processes | ULE System | ULE Stddev | BSD System | BSD Stddev | Faster |
|---|---|---|---|---|---|
| 2 | 434.9 | 2.05 | 408.7 | 1.28 | 4BSD |
| 4 | 499.2 | 1.65 | 465.16 | 2.228 | 4BSD |

Table 7.0 System Time Statistics for Schedulers

Now, it looks fairly easy to update operating system source code because it's possible to check if the source code is acting as it should or not.

## 6. FUTURE WORK

This paper is the first step of OS bug detection system. The supposed plan starts by identifying set of possible behaviors for specific components of operating system thus checking updated code against possible behaviors thus fixing the resulted bugs.

## 7. CONCLUSIONS

This paper introduced fairly easy technique to debug the operating system from both correctness and effectiveness point of views according to some similarities in the behavior of different operating systems.

## 8. ACRONYMS

- OS: Operating System
- ddd: Data Display Debugger
- Windbg: Windows Debugger
- GDB: GNU Project Debugger
- Qemu: Quick EMUlator.
- Stddev: Standard Deviation
- BSD: Berkeley Software Distribution
- ULE: A FreeBSD Task Scheduler
- GPL: GNU General Public License.
- EPL: ECLIPSE PUBLIC LICENSE

## 9. ACKNOWLEDGEMENT

I thank Bruce Lester, PhD at Maharishi University of Management and Wail Al-Kilani, Associate Professor at Ain-Shams University for their support. I have learned a lot from Bruce when I got the opportunity to be teaching assistant in his parallel programming class

## 10. REFERENCES

[1] Matthew A. Telles, and Yuan Hsieh, "The Science of Debugging," The Coriolis Group, ISBN 1-57610-917-8, 2001.
[2] Dmitry Vostokov, "Memory Dump Analysis Anthology," Volume 1, OpenTask, ISBN 978-0-9558328-0-2, 2008.
[3] Andreas Zeller, "Why Programs Fail: A Guide to Systematic Debugging," Morgan Kaufmann, ISBN 1-55860-866-4, 2005.
[4] Ray Kinsella, "Profiling and Debugging the FreeBSD* Kernel," Document Number: 321772-001, 2009.
[5] S.David Ribo, and Clearwater, FL, "How to use the Data Step Debugger," Paper 52-25, (available for download from the author's Web site: HTTP://WWW.JADETEK.COM/JADETECH.HTM)
[6] Norman Matloff, The Art of Debugging with Gdb, DDD, and Eclipse, No Starch Press, ISBN: 9781593271749, 2008

**Author**

Mohamed Farag is post graduate student in Maharishi University of Management in USA. In the first six months of 2012, Mohamed worked as teaching assistant in Maharishi University of Management. He also worked as instructor in Ain Shams University in Egypt during the year 2011. In 2010, Mohamed received Google Summer of Code award and was honored by the scientific community in Menoufia University in Egypt. In addition, Mohamed received "The Best Programming Project" award in Egyptian Universities Summit in 2010 and the same award in 2011. Mohamed has been an active contributor in FreeBSD community since May, 2010 and has led ArabBSD project since June, 2011. In 2012, Mohamed was selected to join the Institute for Computer Sciences, Social Informatics and Telecommunications Engineering "ICST", International Association of Computer Science and Information Technology "IACSIT", Computer Science Teachers Association "CSTA ACM" and Academy & Industry Research Collaboration Center "AIRCC". In September 2012, Mohamed was selected to be reviewer for International Journal of Computer Science and Information Technology "IJCSIT" and program committee in AIRCC. Mohamed has published one paper in IJCSIT in August, 2012.